\documentclass{pasa}

\jid{PASA}
\doi{10.1017/pas.\the\year.xxx}
\jyear{\the\year}

\usepackage[authoryear]{natbib}
\bibpunct{(}{)}{;}{a}{}{,}
\usepackage{aas_macros}
\usepackage{hyperref} 
\hypersetup{colorlinks,citecolor=blue,linkcolor=blue,urlcolor=blue}

\def\aegean{{\sc Aegean}}
\def\robbie{{\sc Robbie}}
\def\fitswarp{{\sc Fits\_Warp}}
\def\model{RISS19}
\def\halpha{{\ensuremath{\mathrm{H}_\alpha}}}

\def\rf{{\ensuremath{r_{\mathrm{F}}}}}
\def\rref{{\ensuremath{r_{\mathrm{ref}}}}}
\def\rdiff{{\ensuremath{r_{\mathrm{diff}}}}}
\def\thetadiff{{\ensuremath{\theta_{\mathrm{diff}}}}}
\def\thetascatt{{\ensuremath{\theta_{\mathrm{scatt}}}}}
\def\thetasrc{{\ensuremath{\theta_{\mathrm{src}}}}}

\title[RISS of Extragalactic Radio Sources]{Refractive Interstellar Scintillation of Extra-galactic Radio Sources I: Expectations}
\author[Hancock et al.]{P. J. Hancock$^1$\thanks{email: Paul.Hancock@curtin.edu.au} and 
E. G. Charlton $^1$ and
J-P. Macquart $^1$ and
N. Hurley-Walker$^1$\\
\affil{$^1$International Centre for Radio Astronomy Research, Curtin University, Bentley, WA 6102, Australia}%
}%

\begin{document}

\begin{abstract}
Surveys for transient and variable phenomena can be confounded by the presence of extrinsic variability such as refractive interstellar scintillation (RISS).
We have developed an all-sky model for RISS which can predict variability on a variety of timescales, survey locations, and observing frequencies.
The model makes use of \halpha{} intensity maps to probe the emission measure along the line of sight, convert this to a scattering measure, and finally a scintillation strength.
The model uses previously developed and long understood physics along with (indirect) measurements of the electron content and distribution within the Milky Way.
We develop a set of expectations that are useful in the planning of future surveys for transient and radio variability, and demonstrate that the $1$-GHz sky is a poor predictor of the variable nature of the $100$-MHz sky.
Interestingly, the correlation between the incidence of variability and Galactic latitude which has been seen at $1$\,GHz, is reversed at $100$\,MHz.

We compare the predictions of our model to a low-frequency radio survey that was conducted with the Murchison Widefield Array, and find good qualitative agreement.
We discuss the implications, current limitations, and future development of the model. 
The model has been implemented in a Python code and is available on GitHub/Zenodo.


\end{abstract}
\begin{keywords}
scintillation -- simulation -- keyword3 -- keyword4 -- keyword5
\end{keywords}
\maketitle%
\section{Introduction}
\label{sec:intro}
Variability offers an insight into the conditions and processes that affect radio sources.
By measuring intrinsic variability we can probe the size of the emitting region (light crossing time arguments), the nature of the emission mechanism (coherent vs incoherent, thermal vs non-thermal), or even the recent history of the object (light curves of Novae mapping to mass loss history).
In turn, extrinsic variability can be used to probe the intervening medium between the observer and the source, and can be used to measure the degree and nature of turbulence in the Milky Way \citep[see][for an overview]{bignall_variable_2005}. 

It was initially assumed at frequencies below $750$\,MHz no intrinsic variability would be detected, and so it was not seriously investigated until \citet{hunstead_four_1972} showed evidence for variability in four radio sources at $408$\,MHz.
In the following years, many more observations were made and the phenomenon of low frequency variability received considerable theoretical attention \citep[see][for a summary]{rickett_radio_1990}.
Variability was observed on timescales as long as years, but also as short as days.
These observed timescales of variability would imply a very small emission region, and a corresponding brightness temperature that exceeds the Compton limit \citep{kellermann_spectra_1969}, arguing against intrinsic variability \citep[although, see][]{wagner_intraday_1995}.
The variability seen by \citet{hunstead_four_1972} is now widely agreed to be due to inter-stellar scintillation (ISS) \citep{rickett_refractive_1986}.

Many studies have noted that radio variability is more common at low Galactic latitudes \citep[eg,][]{cawthorne_low_1985,gaensler_long-term_2000, lovell_micro-arcsecond_2008, bannister_22-yr_2011}.
The interpretation is that the variability is caused by (refractive) ISS and that the dependence on $|b|$ is because the scattering is due to material in our own Galaxy and at low latitudes our line of sight intersects a greater amount of this material.
Indeed \citet{lovell_micro-arcsecond_2008} point out that the incidence of variability in their compact radio sources is more strongly correlated with \halpha{} (a tracer of free electrons) than with $|b|$.

The time investment required to survey for variability is necessarily greater than a single epoch survey of the static sky.
There has thus been a new industry of variability surveys over the last decade or so, that leverage telescope archives to identify regions of sky that were observed multiple times, and use these to form the basis of a survey for variability.
For example \citet{bannister_22-yr_2011} used 3000 partially overlapping observations from the archive of the Molonglo Observatory Synthesis Telescope (MOST) to survey the southern sky for variable and transient events over the 22-year history of the archive at $843$\,MHz.
A large fraction of the MOST archive consists of observations for the Sydney University Molonglo Sky Survey \citep[SUMSS; ][]{mauch_sumss_2003} and the Molonglo Galactic Plane Survey \citep[MGPS; ][]{murphy_second_2007}, and their design means that only a small amount of the total sky is observed more than once.
A similar approach was taken by \citet{thyagarajan_variable_2011} who used overlapping and repeated observations from the Faint Images of the Radio Sky at Twenty-Centimeters
 survey \citep[FIRST; ][]{becker_first_1995} to search for variability at $1.4$\,GHz.
A more focused approach was taken by \citet{bell_automated_2011}, who used archival VLA observations of seven calibrator fields to search for variability at $4.8$ and $8.4$\,GHz.
Since the chosen fields were commonly-observed calibrators, these fields had $100-1000$ epochs available in the archive.
Another approach is to draw from overlapping surveys such as the FIRST survey and a survey of the Sloan Digital Sky Survey (SDSS) stripe \citep{hodge_high-resolution_2011,hodge_millijansky_2013}, or the Murchison Widefield Array (MWA) GLEAM \citep{hurley-walker_galactic_2017} and Tata
Institute for Fundamental Research GMRT Sky Survey (TGSS) Alternative Data Release 1 (ADR1) \citep{intema_gmrt_2017} surveys \citep{murphy_search_2017}.
This {\em ad-hoc} blind survey avoids many of the biases introduced by targeted surveys such as the Micro-Arcsecond Scintillation-Induced Variability \citep[MASIV,][]{jauncey_microarcsecond_2007} survey, but does so at the expense of good survey design.
Not including meaningful transient and variability analysis planning into survey design means that the replication of previous results becomes a matter of chance\footnote{See \citet{hancock_radio_2016} for a fortuitous example of where an archival survey could resolve some of the tension between the reported density of variable sources seen in previous surveys.}.

As new wide-field instruments such as the MWA \citep{tingay_murchison_2013}, the Australian SKA Pathfinder \citep[ASKAP,][]{johnston_science_2008}, and the Square Kilometer Array \citep[SKA,][]{dewdney_square_2009} come online, the required time to survey large amounts of the sky drastically decreases and the feasibility of a pre-planned, large-area, sensitive, blind survey greatly increases \citep{murphy_vast_2013, bell_murchison_2018}.
If large surveys for transient and variable radio sources are to be carried out, it is incumbent on the survey designers to understand the expected outcomes of these surveys in terms of the number of events expected from already known physics, and source populations.

\citet{pietka_variability_2014} present a fairly comprehensive overview of the variable and transient events that are known, as well as their expected brightness and areal density.
From this and similar works it is possible to compute the expected number of variable or transient events in a given survey, or to design a survey to maximize the detection of a desired class of object \citep[eg,][]{fender_radio_2011}.

As well as modelling the expected rates of intrinsic variability it is important to model the extrinsic variability.
The primary source of extrinsic variability is scintillation caused by changes in electron density in the intergalactic, interstellar, or interplanetary medium, and also in the Earth's ionosphere.
Theoretically scintillation is well understood and has been described previously by \citet{rickett_refractive_1986,rickett_radio_1990} and \citet{narayan_physics_1992}.
The degree and nature of scintillation is dependent on the amount of free electrons between the observer and the source, as well as the relative motion of each, and the observing frequency.
Scintillation is not dependent on the temperature of the intervening electrons, and since cold electrons are most abundant, they contribute the most; they are otherwise a difficult population to observe directly.
Luckily, the line-of-sight electron density can be measured by observing the frequency-dependent dispersion and scattering delay of a short pulsed signal such as that from a pulsar \citep{taylor_pulsar_1993}.
Using the observed dispersion measures (DM) of pulsars, and ISS of extra-galactic sources \citet{taylor_pulsar_1993} created a quantitative model of the Galactic free electron distribution.
This quantitative model was accompanied by a set of Fortran sub-routines which implemented the model.
Using this model, \citet{walker_interstellar_1998} presented maps of the transition frequency and scattering disk size to give expectations for the scintillation behaviour of compact extra-galactic radio sources.

The model of \citet{taylor_pulsar_1993} has been updated and refined over the years by \citet{cordes_ne2001.i._2002} and most recently \citet{yao_new_2017}.
These models grew from the pulsar community and have been widely used and accepted within this community, as well as the emerging FRB community.
In both PSR and FRB studies, the primary focus is on diffractive ISS (DISS), as this causes large-amplitude, short-duration variations with a small decorrelation bandwidth.
So whilst refractive ISS (RISS) may still affect pulsar and FRB observations, the effect is largely ignored due to the much smaller variability ($\sim 10\%$), longer timescales (hours to years), and larger decorrelation bandwidth.
Therefore the models of electron density, and the accompanying code which implements them, are largely focused on predicting the the DISS behaviour of ultra-compact Galactic and extra-galactic radio sources.
The utility of these models has not yet found its way into studies of slow variability, in part because the relevant data are not reported directly, and because much of the pulsar population is Galactic, where as many surveys for variability purposely avoid the Galactic plane.
Extra-galactic radio sources are rarely compact enough to exhibit DISS, but can still exhibit RISS.
Since DISS and RISS are caused by the same underlying material, models like that of \citet{yao_new_2017} can be used to predict the degree of RISS expected from a point source.
However, the programs which embed these models do not report RISS statistics directly, and the models themselves are not calibrated against measurements of RISS.
Further more the community of astronomers who search for variability in extra-galactic radio sources can be easily misled by RISS\footnote{For example, \citet{keane_host_2016} claimed to detect a afterglow coincident with an FRB which turned out to be scintillation of a background AGN.}, in part because there are no RISS focused programs that are as easy to access as \citet{cordes_ne2001.i._2002} or \citet{yao_new_2017}.
To this end we endeavour to model the variability induced by RISS, present the theory in a manner that is easily accessible, and embed the model in an easy to use Python program.

Motivated by the work of \citet{lovell_micro-arcsecond_2008}, and a desire to produce an all-sky model, we start with \halpha{} intensity as our input data, and produce an all-sky model of RISS.
We present here a theoretical and practical framework for modelling RISS which can be built upon as improvements are made in the theory and in the input data upon which the model is based.
We begin firstly with an overview of the physics of ISS, and then present our theoretical model.
We then discuss the current limits of our model before making a qualitative comparison with a survey designed to detect radio variability at $185$\,MHz.
We have developed a practical framework which is in the form of a Python code, which was inspired by the NE2001 Galactic Free Electron Density Model\footnote{available at \href{https://www.nrl.navy.mil/rsd/RORF/ne2001/}{www.nrl.navy.mil/rsd/RORF/ne2001}} of \citet{cordes_ne2001.i._2002}
Details of the code, along with the source, can be found on \href{https://github.com/PaulHancock/\model}{GitHub}\footnote{available at \href{https://github.com/PaulHancock/\model}{github.com/PaulHancock/\model}}.

This paper is organized as follows:
In \S\,\ref{sec:ISS} we give an overview of interstellar scintillation with a focus on the strong scattering regime.
In \S\,\ref{sec:modeling} we present a model for computing scintillation parameters based using \halpha{} maps as input data.
In \S\,\ref{sec:expectations} we summarize the model by presenting a set of expectations for observers, both at low ($100$\,MHz) and intermediate ($1$\,GHz) frequencies.
The limitations of our model are discussed in \S\,\ref{sec:limits}.
We discuss implications for future radio surveys in \S\,\ref{sec:implications}.
We make a qualitative comparison between our expectations and a low-frequency radio survey in
\S\,\ref{sec:data}-\ref{sec:comparison}.
Finally, we summarise our work in \S\,\ref{sec:conclusions}.

\section{Interstellar Scintillation}
\label{sec:ISS}
Interstellar Scintillation has been previously described in depth by \citet{narayan_physics_1992, rickett_refractive_1986, rickett_radio_1990}, and \citet{walker_interstellar_1998}.
Here we provide an overview of the relevant results.

Plane waves travel from sources, pass through the ISM, and are then incident upon an observer on Earth.
Since the ISM is a charged plasma, variations in the electron density ($n_e$) will distort the wave-fronts, causing a phase delay or focusing/defocusing.
This effect is interstellar scintillation (ISS) and is the radio analogue of stars twinkling in the night sky due to the Earth's atmosphere.

It is traditional (and easy) to model scintillation using a thin screen model, which assumes that the path integrated effect of the intervening medium can be modeled by a single phase screen at some intermediate distance D.

When considering the effect of the interstellar medium on observations of an extra-galactic source, there are two length scales which are important.
The first is the Fresnel scale \rf{}, which represents the transverse separation over which the phase delay is 1\,rad, and is related to the distance to the phase screen and wavelength via:
\begin{equation}
\rf = \sqrt{\frac{\lambda D}{2\pi}}
\label{eq:rf}
\end{equation}

The second is the diffractive length scale \rdiff{} which represents the transverse separation over which the rms phase variance is 1\,rad.
This scale is dependent on the density variations within the ISM, which are typically modeled by Kolmogorov turbulence. 
\citet{Macquart_temporal_2013} give the following relation (their Eq.\,7a) to compute \rdiff{}:

\begin{equation}
\rdiff = \left[2^{2-\beta}\frac{\pi r^2_e\lambda^2_0\beta}{(1+z_L)^2} SM \frac{\Gamma(-\frac{\beta}{2})}{\Gamma(\frac{\beta}{2})} \right] ^{1/(2-\beta)}
\end{equation}
\noindent where $r_e$ is the classical electron radius, $z_L$ is the red-shift of the scattering material, $\beta$ is the power-law index for the turbulence, and SM is the scattering measure.
We adopt values of $z_L=0$ and $\beta=11/3$ as we assume that scattering material is entirely within our Galaxy, and that the turbulence is Kolmogorov.
With these assumptions the above equation reduces to:
\begin{equation}
\rdiff = 3.7\times 10^9 \left(\frac{\lambda}{1m}\right)^{-6/5} \left(\frac{SM}{10^{12} m^{-17/3}}\right)^{-3/5}\, (\mathrm{m})
\end{equation}

Withe these two length scales in mind, we can now distinguish two scattering regimes of strong and weak scattering.
A useful parameter in this discussion is the scintillation strength $\xi$, defined as:
\begin{equation}
    \xi = \rf/\rdiff
    \label{eq:xi}
\end{equation}

The scaling of $\xi$ with wavelength and scattering measure is:
\begin{equation}
    \xi \propto \lambda^{17/10} \cdot \mathrm{SM}^{3/5}
    \label{eq:xi_prop}
\end{equation}

When $\xi\sim 1$ the scattering is transitional and neither the weak nor strong scintillation discussed below is an adequate description of the relevant physics.
The frequency at which this transition occurs is called the transition frequency with weak scattering occuring above and strong scattering occurring below this frequency.

\subsection{Weak scattering - $\xi \ll 1$}
Weak scattering occurs when the diffractive scale is much larger than the Fresnel scale ($\rdiff \gg \rf$).
In this regime the rms phase variations within the Fresnel zone are small.
Phase variation within the Fresnel zone is dominated by differential path length of nearby light rays, leading to a strong correlation in the phase differences across a large bandwidth.
The scintillation time-scale is related to the relative velocity of the screen and observer, and the fractional variation is low.
Weak scattering is typically observed within our Solar system and gives rise to interplanetary scintillation \citep{hewish_irregular_1955}.
As will be shown in \S~\ref{sec:RISS}, at 1\,GHz or lower, extra-galactic sources are always observed in the strong scattering regime.
Indeed \citet{walker_interstellar_1998, walker_erratum_2001} presents figures of the transition frequency in Galactic coordinates, showing that in order to observe weak scattering, an observing frequency of at least $10$\,GHz is needed over most of the sky.
We will not discuss weak scattering further.

\subsection{Strong scattering - $\xi \gg 1$}
Strong scattering occurs when the diffractive scale is much smaller than the Fresnel scale ($\rdiff \ll \rf$).
In this regime the rms phase variations within the Fresnel zone are large, and the Fresnel scale loses relevance.
Strong scattering is dominated by multipath propagation and gives rise to two different kinds of scintillation: diffractive and refractive.
Diffractive scintillation has fractional flux modulation of unity, is correlated across only a small fractional bandwidth (a few percent), and the variations are rapid in time (seconds to minutes).
Refractive scintillation has lower fractional variation, is correlated across a large bandwidth, and has a lower variation in time.
In both cases a critical angular scale is present that separates sources into being either point like or extended, with the extended sources having a modulation index that is lower and timescale of variability which is longer than the point sources.

\subsubsection{Diffractive scintillation}
\label{sec:DISS}
The multipath propagation of strong scattering results in an observer seeing light scattered from multiple points on the scattering screen.
The individual patches have a size and inter-patch spacing that is approximately \rdiff{}.
Each patch scatters light over an angle of \thetascatt{}:
\begin{equation}
    \thetascatt \sim \rref/D = \frac{1}{k \rdiff}
\end{equation}
\noindent where $k$ is the wave number, and \rref{} is the refractive length scale:
\begin{equation}
    \rref = \rf^2 / \rdiff
\end{equation}
Since the observer sees light arriving from a range of directions within \thetascatt{} of the source, the source will appear to be larger (scatter-broadened) by this same angle.
The critical angular scale for diffractive scintillation is 
\begin{equation}
\thetadiff = \rdiff /D \ll \thetascatt
\end{equation}
The angle \thetadiff{} is so small that only extremely compact sources are able to exhibit diffractive scintillation: pulsars and FRBs being two exemplars of this.
The NE2001 and related models describe and predict the relevant properties for diffractive scintillation at GHz frequencies \citep{cordes_ne2001.i._2002,yao_new_2017}.

\subsubsection{Refractive scintillation}
\label{sec:RISS}
Refractive scintillation is due to large scale inhomogeneities in the interstellar medium with length scales of \rref{}, much larger than in the diffractive case.
Refractive scintillation is a broadband phenomena, and the time-scale for variability is also larger than in the diffractive case:
\begin{equation}
    t_\mathrm{ref} = \rref / v
\end{equation}
The division between point and extended sources (the scattering disk) is now:
\begin{equation}
    \thetascatt = \rref / D \gg \thetadiff
\end{equation}
\noindent so that sources of larger angular scales are able to exhibit refractive interstellar scintillation.
Thus radio galaxies can scintillate due to the interstellar medium, but only in the refractive regime.

The modulation index for a point source is the fractional variation:
\begin{equation}
\begin{split}
    m_p &\equiv  \frac{\sigma}{\mu} \\
    m_p &=  \left( \rdiff/\rf \right)^{1/3}  = \xi^{-1/3} < 1 \label{eq:mp}
\end{split}
\end{equation}
\noindent where $\sigma$ and $\mu$ are the standard deviation and mean flux.
For extended sources (of size \thetasrc{}) the modulation index decreases while the scintillation timescale increases:
\begin{equation}
   m_e = \left( \frac{\rdiff}{\rf}\right)^{1/3} \left(\frac{\thetascatt}{\thetasrc} \right)^{7/6} < m_p < 1
   \label{eq:me}
\end{equation}
   
\begin{equation}  
   t_\mathrm{scint} = t_\mathrm{ref} \left(\frac{\thetasrc}{\thetascatt} \right) = \frac{D\thetasrc}{v}
   \label{eq:tscint}
\end{equation}

\section{Modeling RISS}
\label{sec:modeling}
To model RISS we ultimately need to measure or predict the diffractive scale \rdiff.
Equation \ref{eq:SMshort} shows that with some reasonable assumptions about the ISM, it is possible to relate \rdiff{} to the scattering measure.
SM is a difficult quantity to measure directly, however we can use the emission measure as a proxy.
We use \citet{Haffner_faint_1998}, Eq. 1 and \citet{cordes_ne2001.i._2002} Eq. 16 to convert between $H_\alpha$ intensity ($I_{H_\alpha}$) and scattering measure (SM):

\begin{equation}
\mathrm{SM} = \left(\frac{I_{H_\alpha}}{198R}\right)T_4^{0.9}\frac{\varepsilon^{2}}{(1+\varepsilon^2)}\ell_0^{-2/3}\, (\mathrm{kpc.m}^{-20/3})
\label{eq:SMfull}
\end{equation}

\noindent where $I_{H_\alpha}$ is measured in Rayleighs, $T_4$ is the gas temperature in units of $10^4$\,K, $\varepsilon$ is the fractional variance of $n_e$ inside clouds ($\varepsilon^2 = \langle(\delta n_e)^2\rangle/\bar n^2_e $), and $\ell_0$ is the outer scale of turbulence in units of pc.
We adopt the values of $\varepsilon = 1$,  $T_4=0.8$ \citep{Haffner_faint_1998}, and $\ell_0 = 10^{18}m = 32pc$ \citep{Armstrong_electron_1995}.
With these choices of parameters the resulting equation for SM is then simply a linear rescaling of the  \halpha{} intensity:

\begin{equation}
\mathrm{SM} = 1.26\times 10^{16} I_{H_\alpha}\, (\mathrm{m}^{-17/3})
\label{eq:SMshort}
\end{equation}

\begin{figure}[h]
    \centering
    \includegraphics[width=0.95\linewidth]{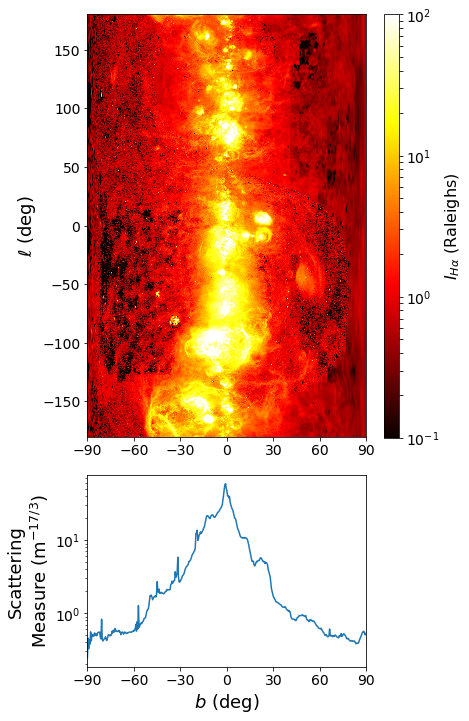}
    \caption{Top: \halpha{} intensity map from \citet{finkbeiner_full-sky_2003}.
    Bottom: Average scattering measure (SM) as a function of galactic latitude, as computed from Eq\,\ref{eq:SMshort} using the \halpha{} intensity map.
    }
    \label{fig:SM}
\end{figure}

\citet{finkbeiner_full-sky_2003} provides an all sky map of \halpha{} by combining data from three different surveys: the Virginia Tech Spectral line Survey \citep[VTSS,][]{dennison_imaging_1998}, the Southern H-Alpha Sky Survey Atlas \citep[SHASSA,][]{gaustad_robotic_2001}, and the Wisconsin H-Alpha Mapper survey \citep[WHAM,][]{reynolds_wisconsin_1998}.
Applying the above equation to the \halpha{} maps of \citet{finkbeiner_full-sky_2003} we can map the SM as a function of viewing direction as in the upper panel of Figure\,\ref{fig:SM}.

With a sky map of SM, and the associated error map derived from the \halpha{} error map, it is then as simple task to calculate Eq\,\ref{eq:rf}-\ref{eq:tscint} over the entire sky.
To this end we have developed a python library \model{}, which carries out these calculations, and a command line tool to interface with this library {\bf ascl/zenodo citation when accepted}.

In order to determine the the scintillation properties along a given line of sight we also need to know the distance to the scattering screen.
A basic model for the scattering screen distance is used in this work: the scattering screen is situated half way between the Earth and the `edge' of the Galaxy.
Our default model consists of a flat disk model of the Milky way, with a radius of 16\,kpc, a height of 1\,kpc, and with the Earth/Sun system sitting 8\,kpc from the Galactic center.

We will now use the modeling software to demonstrate how the SM shown in Figure\,\ref{fig:SM} gives rise to a range of expected (and un-expected) behaviour.

\section{Expectations}
\label{sec:expectations}
\begin{figure}[h]
    \centering
    \includegraphics[width=0.95\linewidth]{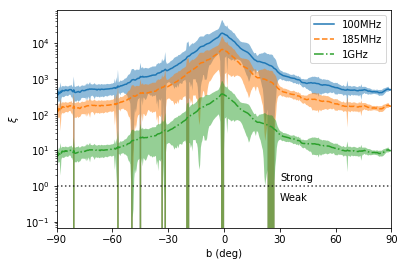}
    \caption{The average scintillation strength ($\xi$) as a function of Galactic latitude.
    The shaded regions represent the $1\sigma$ range of $\xi$ in each $|b|$ bin.}
    \label{fig:xi}
\end{figure}

In this section we take the theory presented in section \ref{sec:ISS} along with the data and code presented in section \ref{sec:modeling}, and develop some evidence-backed expectations of what an observer should expect to see when conducting a survey for variability.
In particular, the effects of frequency, observing direction, and cadence have an interesting interplay that gives rise to some enlightening results.
Many of the relations will be explored with respect to $\xi$ which is shown as a function of Galactic latitude in Figure\,\ref{fig:xi} for the low ($100$\,MHz) and mid ($1$\,GHz) frequency case, as well as $185$\,MHz which corresponds to the survey described in \S\,\ref{sec:data}.
Note that at all frequencies $\xi \gg 1$ so that in all cases we are in the strong scattering regime.

\begin{figure}[h]
    \centering
    \includegraphics[width=0.95\linewidth]{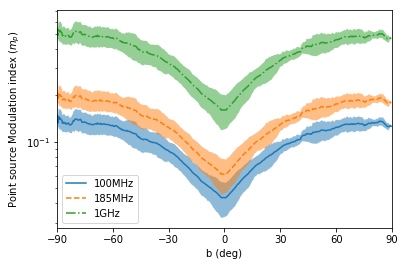}
    \caption{Point source modulation index as a function of Galactic latitude.
    The shaded regions represent the $1\sigma$ range of $m_p$ in each $|b|$ bin.}
    \label{fig:mp}
\end{figure}

The modulation index is related to the scintillation strength by $m_p = \xi^{-1/3}$, and varies with Galactic latitude as shown in Figure\,\ref{fig:mp}.

\subsection*{The effects of source size}
A radio galaxy typically has multiple emission components on a range of different physical scales: a compact ($<$\,pc) nucleus, star-forming regions ($0.1$\,pc - $10$\,kpc), galaxy-wide diffuse emission ($\sim 100$\,kpc), halo and cluster emission ($<1$\,Mpc), as well as jets ($100$\,kpc -- $1$\,Mpc), lobes ($1$\,Mpc), and hot-spots within lobes ($10-100$\,kpc).
Every emission site will be affected by the intervening medium and thus the observed total flux density will vary as a superposition of scintillating components.
When the emission sites have an angular separation less than the scattering disk, \thetascatt{}, their flux density variations will be highly correlated and thus they will be seen to vary together.
Thus the integrated flux density $S$ of a radio galaxy will be distributed between compact and non-compact regions, with the end result that only a fraction of the total flux density will scintillate.

Consider a source with some distribution of emission regions.
If this source were viewed at a Galactic pole, where the scattering disk is small, then some fraction $x$ of the flux will scintillate and a modulation index of $m$ will be observed.
If the source were viewed closer to the Galactic plane, where the scattering disk is larger, then the fraction of flux scintillating will be increased to $x^\prime > x$.
This will increase the observed modulation index.

\begin{figure}[h]
    \centering
    \includegraphics[width=0.95\linewidth]{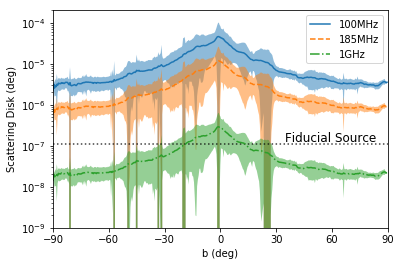}
    \caption{
The scattering disk size at $1$\,GHz (blue, upper), $185$\,MHz (orang, mid), and $100$\,MHz (green, lower).
The shaded regions represent the $1\sigma$ range of $\theta$ in each $|b|$ bin.
A fiducial $10$\,pc source at $1$\,Gpc has an angular size as shown.
When a source subtends an angle larger than the scattering disk it can no longer be modeled as a point source.
    }
    \label{fig:theta}
\end{figure}

There is an additional effect that needs to be considered\footnote{Additionally the intrinsic size of the emission regions in a radio galaxy will change with wavelength, but we ignore that as at most a third-order affect.}: the increased scattering seen towards the plane increases \thetascatt{}, but reduces the intrinsic $m_p$.
The scattering disk scales as:
\begin{equation}
   \thetascatt{} \propto \xi 
\end{equation}
\noindent so that the modulation index scales as:
\begin{equation}
    m_e \propto m_p \thetascatt^{7/6} \propto \xi^{-1/3} \xi^{7/6} = \xi^{5/6} \propto \lambda^{17/12}\cdot\mathrm{SM}^{1/2}
\end{equation}
The net effect is that as a source moves closer to the Galactic plane, where $\xi$ is larger, the effective modulation index increases.
Scattering disk size is shown in Figure\,\ref{fig:theta}.
Thus in a radio survey one should expect the sky density of variable sources to increase towards the Galactic plane, due to RISS.
The claims of \citet{gaensler_long-term_2000, bannister_22-yr_2011, lovell_micro-arcsecond_2008} are all consistent with this expectation.

\subsection*{The effects of timescales}
\begin{figure}[h]
    \centering
    \includegraphics[width=0.95\linewidth]{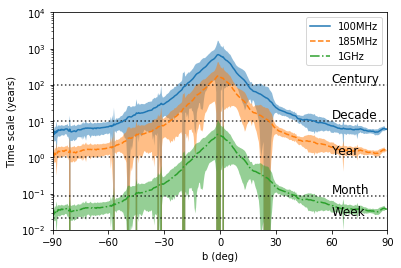}
    \caption{Variability timescale as a function of Galactic latitude.
    The shaded regions represent the $1\sigma$ range of timescales in each $|b|$ bin.
    At $100$\,MHz the timescales are all beyond the reach of typical survey, where as the $1$\,GHz timescales are achievable within a single survey.}
    \label{fig:tscint}
\end{figure}

There is a third parameter that needs to be considered and that is the timescale of variability.
As the scattering becomes stronger, the characteristic timescale for variability will also increase.
The variability timescale scales with frequency as:
\begin{equation}
   t_\mathrm{ref} \propto \nu^{-11/5} 
\end{equation}
The variation of variability timescales is shown in Figure\,\ref{fig:tscint} for our two reference frequencies.

At $1$\,GHz this varies from months to a year depending on whether the sources of interest are towards the Galactic pole or plane respectively.
Since these timescales are comparable to or shorter than the typical observing campaigns for radio variability, the detection of this variability is not affected.
However at $100$\,MHz the variability timescales of decades or longer, depending on the pointing direction.
Thus a low-frequency survey for radio variability with a maximum timescale of $1-5$\,years will show a reduced modulation index.
The modulation index observed when sampling at a cadence of less than the variability time scale varies as:
\begin{equation}
m = m_e \left(\frac{t_\mathrm{obs}}{t_\mathrm{ref}}\right) \propto \xi^{5/6}\xi^{-1} = \xi^{-1/6} \propto \lambda^{-17/60}\cdot\mathrm{SM}^{-1/10}
\label{eq:me_xi}
\end{equation}
Thus at $100$\,MHz the combined effect is to see a weak but negative correlation between the variable source density and absolute Galactic latitude, which is demonstrated in Figure\,\ref{fig:me_t}.

\subsection*{Summary}
\begin{figure}[h]
    \centering
    \includegraphics[width=0.95\linewidth]{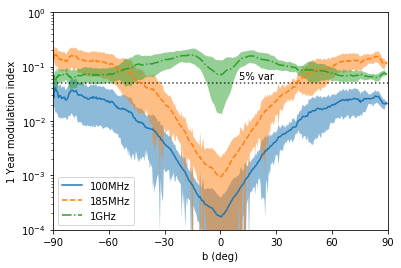}
    \caption{
The expected dependence of modulation index on galactic latitude at $1$\,GHz (green, upper) and $100/185$\,MHz (blue/orange, lower).
The shaded regions represent the $1\sigma$ range of modulation index in each $|b|$ bin.
At $1$\,GHz the modulation index increases toward the Galactic plane, consistent with observations.
However, at $100$\,MHz the modulation index decreases toward the plane, and is an order of magnitude lower than at $1$\,GHz.}
    \label{fig:me_t}
\end{figure}

The arguments outlined in the previous section suggest that we should expect that at $1$\,GHz there is a positive correlation between the sky density of variable sources and absolute Galactic latitude, whilst at $100$\,MHz the correlation reverses due to the timescale of variability being longer than the duration of a typical survey.
Figure\,\ref{fig:me_t} shows the combined effect of frequency, Galactic latitude, source size, and observing timescale on the observed modulation index for compact sources.
The expectations of the previous sections can be seen in this Figure.
Additionally the overall level of variability at $100$\,MHz is at least an order of magnitude less than that at $1$\,GHz.

If we assume that all radio sources contain some compact core and some extended region of emission then we would expect that sources with a larger modulation index would be more likely to be detected as variable, and thus there should be a correlation between the degree of variability and the magnitude of variability.
Therefore we predict that the correlations seen in Figure\,\ref{fig:me_t} should also apply to the areal density of variable sources, with a number of caveats that will be discussed in the next section.

\section{Limitations}
\label{sec:limits}
The model presented in the previous sections is quite detailed, but sadly far from complete.
If the end goal is to compute the expected variability seen in a given survey then there are additional factors that need to be taken into account.
We divide the various factors into three broad categories: assumptions of the theoretical model, limitations of the input data, and limited knowledge of the source population.

\subsection*{Theoretical assumptions}
In our theoretical modeling we made some assumptions which are commonly accepted though difficult to verify.
The fractional variation of $n_e$ inside clouds ($\varepsilon$) and the outer scale of turbulence ($\ell_0$) are critical for converting the observed \halpha{} intensity into a scattering measure, and yet neither can be measured directly.
Indeed these parameters may change with respect to location within the Galaxy but this is not accounted for.

Additionally, our model contains two descriptions of the electron content of the Galaxy, and they are inconsistent.
The \halpha{} map of Figure\,\ref{fig:SM} which is used to derive the SM, implies considerable spatial variation in the electron density, whereas we model the distance to the scattering screen assuming that the Galaxy is cylinder of electrons with uniform density.
Measuring the distance to the scattering screen along multiple lines of sight through the Galaxy is currently prohibitively expensive.
\citet{bower_angular_2013} were able to identify a scattering screen distance of $5.8$\,kpc along a line of sight towards the pulsar SGR\,J$1745-29$, $\approx3''$ from the Galactic Center, by observing the scatter broadening of the pulsar using VLBI.

The 3D structure of the Galaxy, and the electron clouds within, can be probed by pulsar and FRB observations, and leads to empirical models such as \citet{cordes_ne2001.i._2002} and \citet{yao_new_2017}.
Future work for this project is to use the electron distribution models present with these empirical models to produce a new distance model for the scattering screens.
The limitations of such work will be the density of pulsar and FRB observations which are ever increasing, but still much lower than the spatial resolution of the \halpha{} maps.

An alternative approach would be to reformulate the relations in Section\,\ref{sec:ISS} to avoid the explicit use of the screen distance, as has been done for Eq\,\ref{eq:tscint}.
This could be accomplished by using the so-called Bhat relation \citep{bhat_multifrequency_2004} to relate the dispersion measure (DM) to the scattering delay ($\tau_d$).
The DM models of \citet{cordes_ne2001.i._2002} and \citet{yao_new_2017} rely on the same models of Galactic structure, however since the models are calibrated against observed DM, the DM predictions should be more accurate than the (derived) model of the line-of-sight electron distribution.
An approach which avoids the explicit use of distance could therefore be more accurate, however it will still be limited by the relatively low surface density of pulsar and FRB observations.
As telescopes such as the Canadian Hydrogen Intensity Mapping Experiment \citep[ CHIME,][]{collaboration_chime_2018} and ASKAP find an increasing number of FRBs \citep{chime/frb_collaboration_observations_2019, shannon_dispersion-brightness_2018} models of DM and $\tau_d$ will become more precise, making this approach much more attractive.

\subsection*{Observational data}
\begin{figure}
    \centering
    \includegraphics[width=0.9\linewidth]{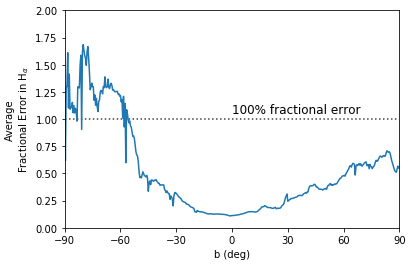}
    \caption{The fractional error in \halpha{} intensity as a function of latitude.}
    \label{fig:ha_err}
\end{figure}

The maps of \citet{finkbeiner_full-sky_2003} are particularly useful for our work in that they provide \halpha{} emission over the entire sky.
However these maps are limited by the input survey data.
In particular, the uncertainty associated with the measurements can be 100\% or greater, particularly in regions of low surface brightness as shown in Figure\,\ref{fig:ha_err}.
Since these uncertainties are propagated through our model, this limitation is apparent to the end user.
Additionally the process of removing the stellar contribution from the input data creates gaps in the data, which are recovered using a smoothing process.
This smoothing process limits the native resolution to $6$\,arcmin.
The smoothing of the data and the low spatial resolution means that the small scale peaks and troughs which contribute to RISS are missed.
The end result is that our model will under-predict the modulation index and over-predict the timescale of variability where small scale structures are present.

Observations of the global \halpha{} emission from a single instrument such as WHAM, would provide a more sensitive and homogeneous input data, though would not improve the spatial resolution.

\subsection*{Source structure}
Radio sources are not all AGN cores, and the fraction of radio emission that is compact vs diffuse varies by source.
A compact source embedded within a diffuse region of emission with a fraction $f$ of the flux density from the compact component, will have an observed modulation index that is reduced by a factor of $1-f$, but the timescale of variability would be unaffected.
Since our model predicts on the behaviour of the compact component it is therefore an upper limit on the true modulation index.

\subsection*{Other sources of extrinsic variability}
The model presented here considers only interstellar scintillation, however at low frequencies ionospheric \citep{loi_power_2015} and interplanetary \citep{morgan_interplanetary_2018} scintillation (IPS) are also present.
The different regimes of scintillation each have different characteristic timescales and spatial scales.
Ionospheric scintillation has a larger \thetascatt{} but shorter timescale than RISS, so that larger sources can exhibit scintillation.
The timescale of variability for ionospheric ($\sim 10$\,sec) and interplanetary ($\sim 1$\,sec) at $100$\,MHz are different enough from the interstellar scintillation that they can be easily separated.
However, minute-long observations spaced by weeks/months/years will still retain some residual of the variability at shorter timescales.
Observations which average over 60\,sec of data will average over $\sim 60$ realizations of interplanetary scintillation, and thus reduce the modulation index by a factor of $\sqrt{60}$.

The observed modulation index will be the quadrature sum of all the various contributions, and thus our model will again under-predict the degree of variability that is observed.
In practice it is possible to form a power spectrum of the brightness fluctuations of a source, and the contribution from the various scattering regimes can be separated based on their differing intrinsic timescales.

Ionospheric and interplanetary scintillation are both in the weak scattering regime even at $100$\,MHz.
Using the MWA \cite{morgan_interplanetary_2018} observe $\sim 10\%$ variability associated with interplanetary scintillation, whilst \cite{loi_power_2015} see $\sim 1\%$ variability associated with ionospheric scintillation.
The typical snap-shot imaging time for the MWA is 2\,min, which means that surveys for variable and transient events such as the MWA Transints Survey \citep[MWATS,][]{bell_murchison_2018}, should expect to see an additional $0.5-1\%$ variability on top of what is predicted for RISS.

\section{Implications for future low-frequency radio surveys}
\label{sec:implications}
The low degree ($\sim 1\%$) and long timescale (years$-$decades) of RISS predicted at low radio frequencies has a few interesting implications.
At low frequencies the line-of-sight electron density can be easily probed with short coherent bursts by measuring a dispersion measure and dispersion delay.
The distribution of electrons, and turbulence within the ISM will be much more difficult to probe at low frequencies due to the small number of sources which will show significant variability.
Surveys for short to intermediate ($<$year) timescale variability should therefore enjoy a very low background contamination from RISS, and be dominated by intrinsic variability.
This adds additional support for SKA\_Low surveys which will focus on rapid follow-up and identification of transient events \citep[eg.][]{fender_transient_2015}.

The differing time-scales of RISS, DISS, IPS, and ionospheric scintillation, mean that low-frequency instruments are preferred for studying the structure of the ionospheric and interplanetary medium.
This has recently been demonstrated in the work of \citet{morgan_interplanetary_2018}.

Low -requency instruments with large fields of view can more confidently rely on so called `in-field' calibration, as the extra-galactic radio sky will be more stable than at higher frequencies.
Such calibration schemes have been used in surveys with the MWA \citep{hurley-walker_galactic_2017, bell_automated_2011}, and consist of using calibrator models which contain hundreds of sources, as opposed to the usual 1--2 bright components used when calibrating instruments with smaller fields-of-view.

\section{HGL survey}
\label{sec:data}
The fact that some of these effects are under-estimates whilst others are upper limits makes it hard to give quantitative predictions.
We will therefore make a qualitative comparison between the model predictions and a low-frequency survey designed to detect RISS.

Not withstanding the limitations mentioned previously, we will compare the expectations of our model with observations from the MWA.

\subsection{Observations}
Observations were taken with the MWA Phase-I layout \citet{tingay_murchison_2013}, under observing program G0003 (PI Hancock).
A total of 9 fields were observed as part of this project, however in this work we report only on the $185$\,MHz observations which were observed as fields 5-8, as described in Table\,\ref{tab:fields}.
A total of 33~epochs of observations were scheduled approximately every week between 2013-08-20 to 2013-12-10 and 2014-08-06 to 2014-12-14, which is when these fields were visible during night time observations.
Of these 33~epochs we were able to obtain calibration solutions and good images for 25.
At the time of observations the MWA primary beam model was not well known.
Each field was scheduled such that it would be observed with the same azimuth and elevation every epoch.
The observations were ``LST-matched''.
The observation details of the 25~calibrated and imaged epochs are shown in Table\,\ref{tab:observations}.

\begin{table}[h]
\centering
\begin{tabular}{cccccc}
Field & RA    & Dec   & $l$     & $b$     & radius\\
      & (deg) & (deg) & (deg) & (deg) & (deg)\\
\hline
5  & 24 & -24 & 198 & -80 & 15 \\
6  & 50 & -20 & 209 & -55 & 15 \\
7  & 65 & -13 & 208 & -39 & 15 \\
8  & 85 & -1  & 205 & -16 & 15 \\
\hline
\end{tabular}
\caption{
The location and size of the fields of interest.
Coordinates are all J2000.
}
\label{tab:fields}
\end{table}

\begin{table*}
\centering
\begin{tabular}{cc|ccccc}
\hline
Epoch & Date & \multicolumn{5}{c}{Observation ID} \\
  & & Field 5    & Field 6    & Field 7    & Field 8    & Calibrator \\
\hline
2  & 2013-08-27 & 1061674440 & 1061674568 & 1061674696 & 1061674824 & 1061673704\\
3  & 2013-09-03 & 1062277584 & 1062277712 & 1062277840 & 1062277968 & 1062276848\\
4  & 2013-09-10 & 1062880736 & 1062880864 & 1062880992 & 1062881120 & 1062880000\\
5  & 2013-09-17 & 1063483880 & 1063484008 & 1063484136 & 1063484264 & 1063483144\\
6  & 2013-10-01 & 1064690184 & 1064690312 & 1064690440 & 1064690568 & 1064689448\\                   
7  & 2013-10-08 & 1065293328 & 1065293456 & 1065293584 & 1065293712 & 1065292592\\                    
9  & 2013-10-29 & 1067102776 & 1067102904 & 1067103032 & 1067103160 & 1067102040\\                    
10 & 2013-11-14 & 1068481400 & 1068481528 & 1068481656 & 1068481784 & 1068480664\\                    
11 & 2013-11-19 & 1068912224 & 1068912352 & 1068912480 & 1068912608 & 1068911488\\                    
12 & 2013-11-26 & 1069515368 & 1069515496 & 1069515624 & 1069515752 & 1069514632\\                    
13 & 2013-12-03 & 1070118520 & 1070118648 & 1070118776 & 1070118904 & 1070117784\\                   
15 & 2014-08-06 & 1091401072 & 1091401200 & 1091401328 & 1091401456 & 1091400336\\                    
16 & 2014-08-12 & 1091918056 & 1091918184 & 1091918312 & 1091918440 & 1091917320\\                    
17 & 2014-08-19 & 1092521208 & 1092521336 & 1092521464 & 1092521592 & 1092520472\\                    
18 & 2014-08-25 & 1093038192 & 1093038320 & 1093038448 & 1093038576 & 1093037456\\                    
19 & 2014-09-02 & 1093727504 & 1093727632 & 1093727760 & 1093727888 & 1093726768\\                    
20 & 2014-09-09 & 1094330648 & 1094330776 & 1094330904 & 1094331032 & 1094329912\\                    
21 & 2014-09-14 & 1094761472 & 1094761600 & 1094761728 & 1094761856 & 1094760736\\                    
22 & 2014-09-23 & 1095536952 & 1095537080 & 1095537208 & 1095537336 & 1095536216\\                    
23 & 2014-09-30 & 1096140096 & 1096140224 & 1096140352 & 1096140480 & 1096139360\\                    
25 & 2014-10-14 & 1097346392 & 1097346520 & 1097346648 & 1097346776 & 1097345656\\                    
26 & 2014-10-24 & 1098208032 & 1098208160 & 1098208296 & 1098208416 & 1098207296\\                    
27 & 2014-11-03 & 1099069680 & 1099069808 & 1099069936 & 1099070064 & 1099068944\\                    
28 & 2014-11-11 & 1099758992 & 1099759120 & 1099759248 & 1099759376 & 1099758256\\                    
29 & 2014-11-17 & 1100275976 & 1100276104 & 1100276232 & 1100276360 & 1100275240\\                
\hline
\end{tabular}
\caption{
Observations used in this work.
For each epoch four fields were observed as well as a single calibrator (Pictor A).
The epoch numbering is as they were observed, and those that were not able to be calibrated and not used in this work are omitted from this table.
}
\label{tab:observations}
\end{table*}

\subsection{Calibration and Imaging}
Calibration and imaging were performed on the NCI Raijin supercomputer using an early version of the GLEAM processing pipeline \citep{hurley-walker_galactic_2017}.
The processing proceeds as follows:

\begin{itemize}
    \item Download data from the MWA archive;
    \item Convert the gpubox files into a measurement set using {\sc Cotter} \citep{offringa_low-frequency_2015};
    \item Use {\sc Calibrate} \citep{offringa_parametrizing_2016} to calibrate using a model of Pictor A;
    \item Apply the calibration solutions to the four target observations;
    \item Use {\sc WSClean} \citep{offringa_wsclean_2014} to create xx, yy polarization images of the target observations; and
    \item Apply an analytic primary beam model to the xx, yy images and combine to form a stokes I image. 
\end{itemize}


The images processed in this way use the MWA `analytic' primary beam model which is now known to be inaccurate.
Since each epoch was observed with the same LST the approximate primary beam model ensures that the relative flux scales are preserved within a field but not between fields.
Using an inaccurate primary beam model with LST matched observations means that the absolute flux scale may be wrong in the outskirts of the images.
However, the relative flux scale is consistent between images and thus no false variable or transients are introduced as a result.
The inaccurate beam model does mean that joining observations from adjacent fields to create mosaicked images is not feasible.

\subsection{Analysis}
We use \robbie{}\footnote{\href{https://github.com/PaulHancock/Robbie}{github.com/PaulHancock/Robbie}} \citep{hancock_robbie_2019} to automate the process of creating light curves for all sources found within each field.
\robbie{} completes the following processing steps:
\begin{enumerate}
\item Correct for ionospheric effects using \fitswarp{}\footnote{\href{https://github.com/nhurleywalker/fits_warp}{github.com/nhurleywalker/fits\_warp}} \citep{hurley-walker_-distorting_2018};
\item Stack all epochs to form a mean image;
\item Find sources in the mean image and measure their fluxes in each epoch using \aegean{}\footnote{\href{https://github.com/PaulHancock/Aegean}{github.com/PaulHancock/Aegean}} \citep{hancock_source_2018}; and
\item Create light curves for all sources calculate variability statistics.
\end{enumerate}

The first stage of the \robbie{} processing ensures that any ionospheric distortions that are present on the scales smaller than the image field of view are corrected for.
For this stage we use the GLEAM catalogue as a position reference.

\robbie{} computes a mean, standard deviation, and $\chi^2$, for each light curve.
Additionally \robbie{} computes the modulation index (m), de-biased modulation index (m$_d$), and a probability of seeing the given light curve given a model of no variability (p\_val).

As is typically done, we classify sources as variable if they have m$_d > 0.05$ and p\_val$<0.001$.
A total of 58 variable sources are found within the four survey regions.
Of these 58 sources, 45 have a match within 1\,arcmin in the {\sc simbad} astronomical database \citep{wenger_simbad_2000}.
These 45 sources are classified as radio galaxies, AGN, or quasars, which are all consistent with the types of sources expected to show RISS.

\section{Comparison of survey results to RISS models}
\label{sec:comparison}

\begin{figure}[h]
    \centering
    \includegraphics[width=0.95\linewidth]{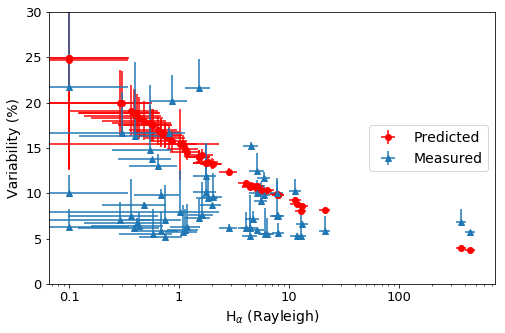}
    \caption{Predicted (red) and measured (blue) degree of variability for sources in the HHL survey as a function of \halpha{} intensity.
    The measurements are the de-biased modulation index (m$_d$) with positive error bars which show the difference between the modulation index and the de-biased modulation index.
    }
    \label{fig:prediction}
\end{figure}

For each of the sources within the survey region we use the \model{} model to predict the modulation index at the location of the source.
We then compare the predicted variability to the measured variability for all sources identified as being variable.
Since the predicted variability is more likely to over-estimate the variability than under-estimate it, we could expect that these predictions act like an envelope function for the measured variability.
The results are shown in Figure\,\ref{fig:prediction}, where the vertical lines around the measured data points represent the difference between the modulation index (m) and the de-biased modulation index (m$_d$), which is a measure of the uncertainty in m$_d$ due to noise.
The predicted variability shows the expected trend of decreasing variability with increasing \halpha{}, which is somewhat followed by the measured variability.
This qualitative agreement between our survey data and model are encouraging and give confidence that with improvements to the model a more accurate prediction could be made.
The fact that the observed variability is larger than the prediction for the highest values of \halpha{} could be an indication that there are small scale enhancements in the electron density in these regions, which are not being captured by our model due to the $6$\,arcmin resolution of the input data.

Some of the limitations of the model, such as the spatial resolution of the input data, mean that a source by source prediction is likely to be beyond our reach.
However, predicting the expected degree of variability within a survey area could be done, if the fraction of compact vs extended sources can be modeled or measured.

\section{Summary and concluding remarks}
\label{sec:conclusions}
We have described a theoretical and empirical approach to predicting the influence of refractive interstellar scintillation (RISS) on our view of the extra-galactic radio sky.
The degree and timescale of variability for a point source is correlated with \halpha{} intensity, and thus weakly correlated with absolute Galactic latitude.
When considering the effects of source size relative to the scattering disk, we show that the correlation between degree of variability and absolute Galactic latitude is inverted at $100$\,MHz compared to that previously seen at $1$\,GHz.
We have implemented our model in a simple Python script which can be used in a way similar to the YMW \citep{yao_new_2017} or NE2001 \citep{cordes_ne2001.i._2002} codes.

Our model has been compared to a survey for variability with the MWA yielding a good qualitative agreement.
Ongoing work (Charlton et al. in prep) will incorporate models of source size and structure, remove the need to model the distance to the scattering screen, and allow for a more quantitative comparison with existing radio surveys.

\begin{acknowledgements}

\subsubsection*{Software}
We acknowledge the work and support of the developers of the following following Python packages: Astropy \citet{the_astropy_collaboration_astropy_2013,astropy_collaboration_astropy_2018}, Numpy \citep{van_der_walt_numpy_2011}, Scipy \citep{Jones_scipy_2001}.
Development of \model{} made extensive use of DS9\footnote{\href{http://ds9.si.edu/site/Home.html}{ds9.si.edu/site/Home.html}} and TOPCAT \citep{Taylor_topcat_2005} for visualization. 
This research made use of Astropy, a community-developed core Python package for Astronomy.

\subsubsection*{Facilities}
This work was supported by resources provided by the Pawsey Supercomputing Centre with funding from the Australian Government and the Government of Western Australia.
This scientific work makes use of the Murchison Radio-astronomy Observatory, operated by CSIRO. We acknowledge the Wajarri Yamatji people as the traditional owners of the Observatory site. 

\subsubsection*{Services}
This research has made use of NASA’s Astrophysics Data System, and the online cosmology calculator tool\footnote{\href{http://www.astro.ucla.edu/~wright/CosmoCalc.html}{www.astro.ucla.edu/$\sim$wright/CosmoCalc.html}} \citet{wright_cosmology_2006}.
This research has made use of the SIMBAD database, operated at CDS, Strasbourg, France \citet{wenger_simbad_2000}.

\end{acknowledgements}

\nocite*{}
\bibliographystyle{pasa-mnras}
\bibliography{HGL_SM2017}

\begin{thebibliography}{}
\makeatletter
\relax
\def\mn@urlcharsother{\let\do\@makeother \do\$\do\&\do\#\do\^\do\_\do\%\do\~}
\definecolor{darkblue}{rgb}{0,0,0.597656}
\def\mndoi{\begingroup\mn@urlcharsother \@ifnextchar [ {\mndoi@} {\mndoi@[]}}
\def\mndoi@[#1]#2{\def\@tempa{#1}\ifx\@tempa\@empty \href
  {http://dx.doi.org/#2} {\textcolor{darkblue}{doi:#2}}\else \href
  {http://dx.doi.org/#2} {\textcolor{darkblue}{#1}}\fi \endgroup}
\def\mn@eprint#1#2{\mn@eprint@#1:#2::\@nil}
\def\mn@eprint@arXiv#1{\href {http://arxiv.org/abs/#1} {{\tt arXiv:#1}}}
\def\mn@eprint@dblp#1{\href {http://dblp.uni-trier.de/rec/bibtex/#1.xml}
  {dblp:#1}}
\def\mn@eprint@#1:#2:#3:#4\@nil{\def\@tempa {#1}\def\@tempb {#2}\def\@tempc
  {#3}\ifx \@tempc \@empty \let \@tempc \@tempb \let \@tempb \@tempa \fi \ifx
  \@tempb \@empty \def\@tempb {arXiv}\fi \@ifundefined
  {mn@eprint@\@tempb}{\@tempb:\@tempc}{\expandafter \expandafter \csname
  mn@eprint@\@tempb\endcsname \expandafter{\@tempc}}}

\bibitem[\protect\citeauthoryear{Armstrong, Rickett  \& Spangler}{Armstrong
  et~al.}{1995}]{Armstrong_electron_1995}
Armstrong J.~W.,  Rickett B.~J.,   Spangler S.~R.,  1995, \mndoi [ApJ] {DOI:
  10.1086/175515}, 443, 209

\bibitem[\protect\citeauthoryear{{Astropy Collaboration} et~al.,}{{Astropy
  Collaboration} et~al.}{2018}]{astropy_collaboration_astropy_2018}
{Astropy Collaboration} et~al., 2018, \mndoi [The Astronomical Journal]
  {10.3847/1538-3881/aabc4f}, 156, 123

\bibitem[\protect\citeauthoryear{Bannister, Murphy, Gaensler, Hunstead  \&
  Chatterjee}{Bannister et~al.}{2011}]{bannister_22-yr_2011}
Bannister K.~W.,  Murphy T.,  Gaensler B.~M.,  Hunstead R.~W.,   Chatterjee S.,
   2011, \mndoi [Monthly Notices of the Royal Astronomical Society]
  {10.1111/j.1365-2966.2010.17938.x}, 412, 634

\bibitem[\protect\citeauthoryear{Banyer, Murphy  \& {VAST
  Collaboration}}{Banyer et~al.}{2012}]{banyer_vast_2012}
Banyer J.,  Murphy T.,   {VAST Collaboration} 2012, ADASS XXI, 461, 725

\bibitem[\protect\citeauthoryear{Becker, White  \& Helfand}{Becker
  et~al.}{1995}]{becker_first_1995}
Becker R.~H.,  White R.~L.,   Helfand D.~J.,  1995, \mndoi [The Astrophysical
  Journal] {10.1086/176166}, 450, 559

\bibitem[\protect\citeauthoryear{Bell et~al.,}{Bell
  et~al.}{2011}]{bell_automated_2011}
Bell M.~E.,  et~al., 2011, \mndoi [Monthly Notices of the Royal Astronomical
  Society] {10.1111/j.1365-2966.2011.18631.x;}, 415, 2

\bibitem[\protect\citeauthoryear{Bell et~al.,}{Bell
  et~al.}{2018}]{bell_murchison_2018}
Bell M.~E.,  et~al., 2018, \mndoi [Monthly Notices of the Royal Astronomical
  Society] {10.1093/mnras/sty2801}

\bibitem[\protect\citeauthoryear{Bhat, Cordes, Camilo, Nice  \& Lorimer}{Bhat
  et~al.}{2004}]{bhat_multifrequency_2004}
Bhat N. D.~R.,  Cordes J.~M.,  Camilo F.,  Nice D.~J.,   Lorimer D.~R.,  2004,
  \mndoi [The Astrophysical Journal] {10.1086/382680}, 605, 759

\bibitem[\protect\citeauthoryear{Bignall, {de Bruyn}  \& Jauncey}{Bignall
  et~al.}{2005}]{bignall_variable_2005}
Bignall H.~E.,  {de Bruyn} A.~G.,   Jauncey D.~L.,  2005, in {{EAS Publications
  Series}}. pp 157--176, \mndoi{10.1051/eas:2005151}

\bibitem[\protect\citeauthoryear{Bower et~al.,}{Bower
  et~al.}{2013}]{bower_angular_2013}
Bower G.~C.,  et~al., 2013, \mndoi [The Astrophysical Journal]
  {10.1088/2041-8205/780/1/L2}, 780, L2

\bibitem[\protect\citeauthoryear{{CHIME/FRB Collaboration} et~al.,}{{CHIME/FRB
  Collaboration} et~al.}{2019}]{chime/frb_collaboration_observations_2019}
{CHIME/FRB Collaboration} et~al., 2019, \mndoi [Nature]
  {10.1038/s41586-018-0867-7}, 566, 230

\bibitem[\protect\citeauthoryear{Cawthorne \& Rickett}{Cawthorne \&
  Rickett}{1985}]{cawthorne_low_1985}
Cawthorne T.~V.,  Rickett B.~J.,  1985, \mndoi [Nature] {10.1038/315040a0},
  315, 40

\bibitem[\protect\citeauthoryear{Collaboration et~al.,}{Collaboration
  et~al.}{2018}]{collaboration_chime_2018}
Collaboration C.,  et~al., 2018, \mndoi [The Astrophysical Journal]
  {10.3847/1538-4357/aad188}, 863, 48

\bibitem[\protect\citeauthoryear{Cordes \& Lazio}{Cordes \&
  Lazio}{2002}]{cordes_ne2001.i._2002}
Cordes J.~M.,  Lazio T. J.~W.,  2002, eprint arXiv:astro-ph/0207156

\bibitem[\protect\citeauthoryear{Dennison, Simonetti  \& Topasna}{Dennison
  et~al.}{1998}]{dennison_imaging_1998}
Dennison B.,  Simonetti J.~H.,   Topasna G.~A.,  1998, \mndoi [Publications of
  the Astronomical Society of Australia] {10.1071/AS98147}, 15, 147

\bibitem[\protect\citeauthoryear{Dewdney, Hall, Schilizzi  \& Lazio}{Dewdney
  et~al.}{2009}]{dewdney_square_2009}
Dewdney P.~E.,  Hall P.~J.,  Schilizzi R.~T.,   Lazio T. J. L.~W.,  2009,
  \mndoi [IEEE Proceedings] {10.1109/JPROC.2009.2021005}, 97, 1482

\bibitem[\protect\citeauthoryear{Fender \& Bell}{Fender \&
  Bell}{2011}]{fender_radio_2011}
Fender R.~P.,  Bell M.~E.,  2011, Bulletin of the Astronomical Society of
  India, 39, 315

\bibitem[\protect\citeauthoryear{Fender, Stewart, Macquart, Donnarumma, Murphy,
  Deller, Paragi  \& Chatterjee}{Fender et~al.}{2015}]{fender_transient_2015}
Fender R.,  Stewart A.,  Macquart J.~P.,  Donnarumma I.,  Murphy T.,  Deller
  A.,  Paragi Z.,   Chatterjee S.,  2015, Advancing Astrophysics with the
  Square Kilometre Array (AASKA14), p.~51

\bibitem[\protect\citeauthoryear{Finkbeiner}{Finkbeiner}{2003}]{finkbeiner_full-sky_2003}
Finkbeiner D.~P.,  2003, \mndoi [The Astrophysical Journal Supplement Series]
  {10.1086/374411}, 146, 407

\bibitem[\protect\citeauthoryear{Gaensler \& Hunstead}{Gaensler \&
  Hunstead}{2000}]{gaensler_long-term_2000}
Gaensler B.~M.,  Hunstead R.~W.,  2000, Publications Astronomical Society of
  Australia, 17, 72

\bibitem[\protect\citeauthoryear{Gaustad, McCullough, Rosing  \&
  Van~Buren}{Gaustad et~al.}{2001}]{gaustad_robotic_2001}
Gaustad J.~E.,  McCullough P.~R.,  Rosing W.,   Van~Buren D.,  2001, \mndoi
  [Publications of the Astronomical Society of the Pacific] {10.1086/323969},
  113, 1326

\bibitem[\protect\citeauthoryear{Haffner, Reynolds  \& Tufte}{Haffner
  et~al.}{1998}]{Haffner_faint_1998}
Haffner L.~M.,  Reynolds R.~J.,   Tufte S.~L.,  1998, \mndoi [The Astrophysical
  Journal] {10.1086/311449}, 501, L83

\bibitem[\protect\citeauthoryear{Hancock, Murphy, Gaensler, Hopkins  \&
  Curran}{Hancock et~al.}{2012}]{hancock_compact_2012}
Hancock P.~J.,  Murphy T.,  Gaensler B.~M.,  Hopkins A.,   Curran J.~R.,  2012,
  \mndoi [Monthly Notices of the Royal Astronomical Society]
  {10.1111/j.1365-2966.2012.20768.x}, 422, 1812

\bibitem[\protect\citeauthoryear{Hancock, Drury, Bell, Murphy  \&
  Gaensler}{Hancock et~al.}{2016}]{hancock_radio_2016}
Hancock P.~J.,  Drury J.~A.,  Bell M.~E.,  Murphy T.,   Gaensler B.~M.,  2016,
  \mndoi [Monthly Notices of the Royal Astronomical Society]
  {10.1093/mnras/stw1486}, 461, 3314

\bibitem[\protect\citeauthoryear{Hancock, Trott  \& {Hurley-Walker}}{Hancock
  et~al.}{2018}]{hancock_source_2018}
Hancock P.~J.,  Trott C.~M.,   {Hurley-Walker} N.,  2018, \mndoi [Publications
  of the Astronomical Society of Australia] {10.1017/pasa.2018.3}, 35

\bibitem[\protect\citeauthoryear{Hancock, {Hurley-Walker}  \& White}{Hancock
  et~al.}{2019}]{hancock_robbie_2019}
Hancock P.~J.,  {Hurley-Walker} N.,   White T.~E.,  2019, \mndoi [Astronomy and
  Computing] {10.1016/j.ascom.2019.02.004}, 27, 23

\bibitem[\protect\citeauthoryear{Hewish}{Hewish}{1955}]{hewish_irregular_1955}
Hewish A.,  1955, \mndoi [Proceedings of the Royal Society of London Series A]
  {10.1098/rspa.1955.0046}, 228, 238

\bibitem[\protect\citeauthoryear{Hodge, Becker, White, Richards  \&
  Zeimann}{Hodge et~al.}{2011}]{hodge_high-resolution_2011}
Hodge J.~A.,  Becker R.~H.,  White R.~L.,  Richards G.~T.,   Zeimann G.~R.,
  2011, \mndoi [The Astronomical Journal] {10.1088/0004-6256/142/1/3}, 142, 3

\bibitem[\protect\citeauthoryear{Hodge, Becker, White  \& Richards}{Hodge
  et~al.}{2013}]{hodge_millijansky_2013}
Hodge J.~A.,  Becker R.~H.,  White R.~L.,   Richards G.,  2013, \mndoi [The
  Astrophysical Journal] {10.1088/0004-637X/769/2/125}, 769, 125

\bibitem[\protect\citeauthoryear{Hunstead}{Hunstead}{1972}]{hunstead_four_1972}
Hunstead R.,  1972, Astrophysical Letters, 12, 193

\bibitem[\protect\citeauthoryear{{Hurley-Walker} \& Hancock}{{Hurley-Walker} \&
  Hancock}{2018}]{hurley-walker_-distorting_2018}
{Hurley-Walker} N.,  Hancock P.~J.,  2018, \mndoi [Astronomy and Computing]
  {10.1016/j.ascom.2018.08.006}, 25, 94

\bibitem[\protect\citeauthoryear{{Hurley-Walker} et~al.,}{{Hurley-Walker}
  et~al.}{2017}]{hurley-walker_galactic_2017}
{Hurley-Walker} N.,  et~al., 2017, \mndoi [MNRAS] {10.1093/mnras/stw2337}, 464,
  1146

\bibitem[\protect\citeauthoryear{Intema, Jagannathan, Mooley  \& Frail}{Intema
  et~al.}{2017}]{intema_gmrt_2017}
Intema H.~T.,  Jagannathan P.,  Mooley K.~P.,   Frail D.~A.,  2017, \mndoi
  [Astronomy and Astrophysics] {10.1051/0004-6361/201628536}, 598, A78

\bibitem[\protect\citeauthoryear{Jauncey et~al.,}{Jauncey
  et~al.}{2007}]{jauncey_microarcsecond_2007}
Jauncey D.~L.,  et~al., 2007, \mndoi [Astronomical and Astrophysical
  Transactions] {10.1080/10556790701596846}, 26, 575

\bibitem[\protect\citeauthoryear{Johnston et~al.,}{Johnston
  et~al.}{2008}]{johnston_science_2008}
Johnston S.,  et~al., 2008, \mndoi [Experimental Astronomy]
  {10.1007/s10686-008-9124-7}, 22, 151

\bibitem[\protect\citeauthoryear{Jones, Oliphant, Peterson  \& {Others}}{Jones
  et~al.}{2001}]{Jones_scipy_2001}
Jones E.,  Oliphant T.,  Peterson P.,   {Others} 2001, {{SciPy}}: {{Open}}
  Source Scientific Tools for {{Python}}

\bibitem[\protect\citeauthoryear{Keane et~al.,}{Keane
  et~al.}{2016}]{keane_host_2016}
Keane E.~F.,  et~al., 2016, \mndoi [Nature] {10.1038/nature17140}, 530, 453

\bibitem[\protect\citeauthoryear{Kellermann \& {Pauliny-Toth}}{Kellermann \&
  {Pauliny-Toth}}{1969}]{kellermann_spectra_1969}
Kellermann K.~I.,  {Pauliny-Toth} I. I.~K.,  1969, \mndoi [The Astrophysical
  Journal] {10.1086/180305}, 155, L71

\bibitem[\protect\citeauthoryear{Lewandowski, Ro{\.z}ko, Kijak, Bhattacharyya
  \& Roy}{Lewandowski et~al.}{2015}]{lewandowski_study_2015}
Lewandowski W.,  Ro{\.z}ko K.,  Kijak J.,  Bhattacharyya B.,   Roy J.,  2015,
  \mndoi [Monthly Notices of the Royal Astronomical Society]
  {10.1093/mnras/stv2159}, 454, 2517

\bibitem[\protect\citeauthoryear{Loi et~al.,}{Loi
  et~al.}{2015}]{loi_power_2015}
Loi S.~T.,  et~al., 2015, \mndoi [Radio Science] {10.1002/2015RS005711}, 50,
  574

\bibitem[\protect\citeauthoryear{Lovell et~al.,}{Lovell
  et~al.}{2008}]{lovell_micro-arcsecond_2008}
Lovell J. E.~J.,  et~al., 2008, \mndoi [The Astrophysical Journal] {DOI:
  10.1086/592485; eprintid: arXiv:0808.1140}, 689, 108

\bibitem[\protect\citeauthoryear{Macquart \& Koay}{Macquart \&
  Koay}{2013}]{Macquart_temporal_2013}
Macquart J.-P.,  Koay J.-Y.,  2013, \mndoi [The Astrophysical Journal]
  {10.1088/0004-637X/776/2/125}, 776

\bibitem[\protect\citeauthoryear{Mauch, Murphy, Buttery, Curran, Hunstead,
  Piestrzynski, Robertson  \& Sadler}{Mauch et~al.}{2003}]{mauch_sumss_2003}
Mauch T.,  Murphy T.,  Buttery H.,  Curran J.,  Hunstead R.,  Piestrzynski B.,
  Robertson J.,   Sadler E.,  2003, Monthly Notices of the Royal Astronomical
  Society, 342, 1117

\bibitem[\protect\citeauthoryear{Morgan et~al.,}{Morgan
  et~al.}{2018}]{morgan_interplanetary_2018}
Morgan J.~S.,  et~al., 2018, \mndoi [Monthly Notices of the Royal Astronomical
  Society] {10.1093/mnras/stx2284}, 473, 2965

\bibitem[\protect\citeauthoryear{Murphy, Mauch, Green, Hunstead, Piestrzynska,
  Kels  \& Sztajer}{Murphy et~al.}{2007}]{murphy_second_2007}
Murphy T.,  Mauch T.,  Green A.,  Hunstead R.~W.,  Piestrzynska B.,  Kels
  A.~P.,   Sztajer P.,  2007, Monthly Notices of the Royal Astronomical
  Society, 382, 382

\bibitem[\protect\citeauthoryear{Murphy et~al.,}{Murphy
  et~al.}{2013}]{murphy_vast_2013}
Murphy T.,  et~al., 2013, \mndoi [PASA] {10.1017/pasa.2012.006;}, 30, 6

\bibitem[\protect\citeauthoryear{Murphy et~al.,}{Murphy
  et~al.}{2017}]{murphy_search_2017}
Murphy T.,  et~al., 2017, \mndoi [Monthly Notices of the Royal Astronomical
  Society] {10.1093/mnras/stw3087}, 466, 1944

\bibitem[\protect\citeauthoryear{Narayan}{Narayan}{1992}]{narayan_physics_1992}
Narayan R.,  1992, \mndoi [Philosophical Transactions: Physical Sciences and
  Engineering, Volume 341, Issue 1660, pp. 151-165] {10.1098/RSTA.1992.0090},
  341, 151

\bibitem[\protect\citeauthoryear{Offringa et~al.,}{Offringa
  et~al.}{2014}]{offringa_wsclean_2014}
Offringa A.~R.,  et~al., 2014, \mndoi [Monthly Notices of the Royal
  Astronomical Society] {10.1093/mnras/stu1368}, 444, 606

\bibitem[\protect\citeauthoryear{Offringa et~al.,}{Offringa
  et~al.}{2015}]{offringa_low-frequency_2015}
Offringa A.,  et~al., 2015, \mndoi [Publications of the Astronomical Society of
  Australia] {10.1017/pasa.2015.7}, 32

\bibitem[\protect\citeauthoryear{Offringa et~al.,}{Offringa
  et~al.}{2016}]{offringa_parametrizing_2016}
Offringa A.~R.,  et~al., 2016, \mndoi [Monthly Notices of the Royal
  Astronomical Society] {10.1093/mnras/stw310}, 458, 1057

\bibitem[\protect\citeauthoryear{Pietka, Fender  \& Keane}{Pietka
  et~al.}{2014}]{pietka_variability_2014}
Pietka M.,  Fender R.~P.,   Keane E.~F.,  2014, \mndoi [Monthly Notices of the
  Royal Astronomical Society, Volume 446, Issue 4, p.3687-3696]
  {10.1093/mnras/stu2335}, 446, 3687

\bibitem[\protect\citeauthoryear{Reynolds, Tufte, Haffner, Jaehnig  \&
  Percival}{Reynolds et~al.}{1998}]{reynolds_wisconsin_1998}
Reynolds R.~J.,  Tufte S.~L.,  Haffner L.~M.,  Jaehnig K.,   Percival J.~W.,
  1998, \mndoi [Publications of the Astronomical Society of Australia] {DOI:
  10.1071/AS98014}, 15, 14

\bibitem[\protect\citeauthoryear{Rickett}{Rickett}{1986}]{rickett_refractive_1986}
Rickett B.~J.,  1986, \mndoi [The Astrophysical Journal] {10.1086/164444}, 307,
  564

\bibitem[\protect\citeauthoryear{Rickett}{Rickett}{1990}]{rickett_radio_1990}
Rickett B.~J.,  1990, \mndoi [ARA\&A] {10.1146/annurev.aa.28.090190.003021},
  28, 561

\bibitem[\protect\citeauthoryear{Schlegel, Finkbeiner  \& Davis}{Schlegel
  et~al.}{1998}]{schlegel_maps_1998}
Schlegel D.~J.,  Finkbeiner D.~P.,   Davis M.,  1998, \mndoi [The Astrophysical
  Journal] {10.1086/305772}, 500, 525

\bibitem[\protect\citeauthoryear{Shannon et~al.,}{Shannon
  et~al.}{2018}]{shannon_dispersion-brightness_2018}
Shannon R.~M.,  et~al., 2018, \mndoi [Nature] {10.1038/s41586-018-0588-y}, 562,
  386

\bibitem[\protect\citeauthoryear{Taylor}{Taylor}{2005}]{Taylor_topcat_2005}
Taylor M.,  2005, in Shopbell P.,  Britton M.,   Ebert R.,  eds,  Astronomical
  {{Society}} of the {{Pacific Conference Series}} Vol. 347, Astronomical
  {{Data Analysis Software}} and {{Systems XIV}}. pp 29--29

\bibitem[\protect\citeauthoryear{Taylor \& Cordes}{Taylor \&
  Cordes}{1993}]{taylor_pulsar_1993}
Taylor J.~H.,  Cordes J.~M.,  1993, \mndoi [The Astrophysical Journal]
  {10.1086/172870}, 411, 674

\bibitem[\protect\citeauthoryear{{The Astropy Collaboration} et~al.,}{{The
  Astropy Collaboration} et~al.}{2013}]{the_astropy_collaboration_astropy_2013}
{The Astropy Collaboration} et~al., 2013, \mndoi [Astronomy \& Astrophysics]
  {10.1051/0004-6361/201322068}, 558, 9

\bibitem[\protect\citeauthoryear{Thyagarajan, Helfand, White  \&
  Becker}{Thyagarajan et~al.}{2011}]{thyagarajan_variable_2011}
Thyagarajan N.,  Helfand D.~J.,  White R.~L.,   Becker R.~H.,  2011, \mndoi
  [The Astrophysical Journal] {10.1088/0004-637X/742/1/49}, 742, 49

\bibitem[\protect\citeauthoryear{Tingay et~al.,}{Tingay
  et~al.}{2013}]{tingay_murchison_2013}
Tingay S.~J.,  et~al., 2013, \mndoi [PASA] {10.1017/pasa.2012.007}, 30, 21

\bibitem[\protect\citeauthoryear{Wagner \& Witzel}{Wagner \&
  Witzel}{1995}]{wagner_intraday_1995}
Wagner S.~J.,  Witzel A.,  1995, \mndoi [Annual Review of Astronomy and
  Astrophysics] {10.1146/annurev.aa.33.090195.001115}, 33, 163

\bibitem[\protect\citeauthoryear{Walker}{Walker}{1998}]{walker_interstellar_1998}
Walker M.~A.,  1998, \mndoi [Monthly Notices of the Royal Astronomical Society]
  {DOI: 10.1046/j.1365-8711.1998.01238.x}, 294, 307

\bibitem[\protect\citeauthoryear{Walker}{Walker}{2001}]{walker_erratum_2001}
Walker M.~A.,  2001, \mndoi [Monthly Notices of the Royal Astronomical Society]
  {10.1046/j.1365-8711.2001.04104.x}, 321, 176

\bibitem[\protect\citeauthoryear{Wenger et~al.,}{Wenger
  et~al.}{2000}]{wenger_simbad_2000}
Wenger M.,  et~al., 2000, \mndoi [Astronomy and Astrophysics Supplement Series]
  {10.1051/aas:2000332}, 143, 9

\bibitem[\protect\citeauthoryear{Wright}{Wright}{2006}]{wright_cosmology_2006}
Wright E.~L.,  2006, \mndoi [Publications of the Astronomical Society of the
  Pacific] {10.1086/510102}, 118, 1711

\bibitem[\protect\citeauthoryear{Yao, Manchester  \& Wang}{Yao
  et~al.}{2017}]{yao_new_2017}
Yao J.~M.,  Manchester R.~N.,   Wang N.,  2017, \mndoi [The Astrophysical
  Journal] {10.3847/1538-4357/835/1/29}, 835, 29

\bibitem[\protect\citeauthoryear{{van der Walt}, Colbert  \& Varoquaux}{{van
  der Walt} et~al.}{2011}]{van_der_walt_numpy_2011}
{van der Walt} S.,  Colbert S.~C.,   Varoquaux G.,  2011, \mndoi [Computing in
  Science \& Engineering] {10.1109/MCSE.2011.37}, 13, 22

\makeatother
\end{thebibliography}

\end{document}